\documentclass[12pt]{article}
\pdfoutput =1
\usepackage{graphics}
\usepackage{graphicx}
\DeclareGraphicsExtensions{.pdf}
\usepackage{float} 
\usepackage{subfloat}
\usepackage{amsmath}
\usepackage{slashed}
\usepackage{amsfonts}  
\usepackage{amssymb}

\usepackage{color}

\usepackage{subcaption}

\begin{document}
\date{}
\title{{\bf{\Large Phases of Euclidean wormholes in JT gravity}}}
\author{
 {\bf {\normalsize Hemant Rathi\thanks{E-mail:  hrathi07@gmail.com, hrathi@ph.iitr.ac.in}~ and ~Dibakar Roychowdhury}$
$\thanks{E-mail:  dibakarphys@gmail.com, dibakar.roychowdhury@ph.iitr.ac.in}}\\
 {\normalsize  Department of Physics, Indian Institute of Technology Roorkee,}\\
  {\normalsize Roorkee 247667, Uttarakhand, India}
\\[0.3cm]
}
\maketitle
\abstract{We present a JT gravity set up that reveals the evidence of (Euclidean) wormhole to black hole phase transition at finite charge density and/or chemical potential. We identify the low temperature phase of the system as the charged
wormhole solution. As the temperature of the system is increased, it
undergoes a first order phase transition to a two black hole system at
finite charge density. At the critical point ($T = T_0$) of the phase transition,
both the Free energy (density) and the charge undergoes a discontinuous
change. Finally, we conjectured that the field theory dual to this
gravitational set up is a two-site (uncoupled) complex SYK model at a
finite chemical potential.}
\section{Introduction and summary}
Wormholes are the geometrical bridges that connect the asymptotic regions of space-time \cite{Hawking:1988ae}-\cite{Hawking:1990in}. These are the solutions of Einstein's equation in the classical limit. 

Wormhole solutions \cite{Maldacena:2018lmt}-\cite{Garcia-Garcia:2019poj} are studied extensively in the context of Jackiw-Teitelboim (JT) gravity models \cite{Jackiw:1984je}-\cite{Almheiri:2014cka} those are conjectured to be dual to Sachdev-Ye-Kitaev (SYK) like models \cite{Sachdev:1992fk}-\cite{Bulycheva:2017uqj} in one dimension. In particular, the authors in \cite{Maldacena:2018lmt} investigate a two site coupled SYK model, which for small values of the coupling, is found to exhibit a \emph{gapped} phase at sufficiently low energies. 

This \emph{gapped} phase in the coupled SYK model is identified with the traversable wormhole solution of the nearly-$AdS_2$ \cite{Almheiri:2014cka}, \cite{Jensen:2016pah} gravity interacting with the matter fields. However, as the temperature is increased, the SYK model exhibits a phase transition which in the language of the dual gravity picture, corresponds to a Hawking-Page transition into a black hole phase at high temperatures.

The authors in \cite{Garcia-Garcia:2020ttf} further extend these results to explore a two site uncoupled Majorana SYK model which also exhibits a \emph{gapped} phase at low temperatures. The gravitational analogue of this phenomenon is proposed to be an Euclidean wormhole solution of JT gravity in the presence of matter couplings.

These results were further generalised in \cite{Garcia-Garcia:2020vyr}, where the authors consider a weak coupling between the two site complex SYK whose gravitational dual corresponds to a traversable wormhole solution \cite{Maldacena:2018lmt} at zero charge density. Using the Schwinger-Dyson equation, they further establish the onset of a first order phase transition in which the wormhole phase transits into a two black hole system at high temperatures\footnote{For higher dimensional wormholes  and the associated phase structure see \cite{Maldacena:2004rf}-\cite{Kundu:2021nwp}.}.

Recently, the complex SYK model has been further investigated in the presence of different chemical potentials \cite{Zhang:2020szi}. At low energies, the authors in \cite{Zhang:2020szi} identify the ground state of the system as an eternal traversable wormhole that connects the two sides at low (averaged) chemical potential. Interestingly, these wormhole solutions transit into a two black hole system at high chemical potential.  

Given the above review on the literature, the purpose of the present work is to initiate a systematic investigation of the Euclidean wormhole solutions at finite (charge) density and in particular, to explore the associated phase stability of the solution at low temperatures. 

Below, we outline the key findings of our analysis.

In the present work, we cook up a theory of Einstein-Maxwell-dilaton (EMD) gravity within the JT gravity framework that exhibits a first order phase transition between the charged wormhole solutions at low temperatures and black hole solutions at high temperatures and fixed chemical potential. In particular, we explore the thermal properties of both of these solutions. We observe that the regularised Free energy density $(\mathcal{F}_{(wh)}^{reg})$ of the wormhole configuration is almost constant at sufficiently low temperatures ($T<T_0$) indicating the presence of ``gapped'' phase in the dual (conjectured) two site complex SYK model \cite{Garcia-Garcia:2020ttf}, \cite{Garcia-Garcia:2020vyr}.

The organization for the rest of the paper is as follows.

$\bullet$ In Section \ref{grsetup}, we emphasize on the role of the Maxwell-Chern-Simons (MCS) term \cite{Deser:1981wh}-\cite{VanMechelen:2019ebr} that appears in topological gauge theories. When coupled to $ AdS_3 $ gravity, following a suitable dimensional reduction (see Appendix \ref{dimreduction}), this results into a dynamical JT gravity model which exhibits a wormhole to black hole phase transition at fixed chemical potential. 

We also carry out a first principle derivation of the quantum stress-energy tensor for the $U(1)$ gauge fields ($A_{\mu}$) that takes into account the double trumpet background. Since gauge fields in 2D are non conformal, therefore the present derivation is significantly different from those of the earlier results reported in \cite{Garcia-Garcia:2020ttf}. 

$\bullet$ In Section \ref{thermoref2}, we carry out a detailed analysis on various thermodynamic entities pertaining to the wormhole as well as the black hole phase. These include computing entities like the ``boundary'' Free energy density ($\mathcal{F}$), total charge ($Q$), temperature ($T$) and the chemical potential ($\mu$). In particular, we express these quantities as, $\mathcal{F}=\mathcal{F}(T,\mu)$ and $Q=Q(T,\mu)$, where both $T$ and $\mu$ are treated as independent parameters of the system.

$\bullet$ In section \ref{secphasetrans}, We explore the variations of $\mathcal{F}$ and $Q$ with temperature $(T)$ while keeping the chemical potential $(\mu=\mu_0)$ fixed. Our analysis reveals that the wormhole phase at low temperature ($T<T_0$) undergoes a first order phase transition (at $T=T_0$) into a two black hole system at finite charge $(Q)$.

$\bullet$ In Section \ref{npbnewsecconject}, we qualitatively  discuss the structure of dual field theory (complex SYK model) associated with the 2D Einstein-Maxwell-dilaton gravity. 

Finally, we conclude in Section \ref{secconc} and discuss future extensions of the present work. 
\section{JT gravity set up in 2D }\label{grsetup}
We begin by considering the following Einstein-Maxwell-dilation (EMD) gravity\footnote{In the Appendix \ref{dimreduction}, we show how the first integral on the R.H.S. of (\ref{action}) appears as a result of dimensional reduction. The second term ($ \sim \int (\partial \chi)^2 $), on the other hand, has been added by hand following the same spirit as that of \cite{Garcia-Garcia:2020ttf}.} in 2D
\begin{align}\label{action}
    S_{JT}=&\int_{\mathcal{M}} d^2x\sqrt{-g}\Big[\Phi(R+2)+a_1\Phi^{2}F^2+a_2\Phi\varepsilon^{\mu\nu}F_{\mu\nu}\Big]+\int_{\mathcal{M}} d^2 x\sqrt{-g}(\partial\chi)^2+\nonumber\\
    &\int_{\partial \mathcal{M}} d \tau \sqrt{-\gamma}\Phi2K+S_{ct},
\end{align}
where $a_1,a_2$ are some arbitrary coupling constants of the theory, $\Phi$ is the dilaton and $\chi$ is a scalar field. Here, $S_{ct}$ is the counter term that is introduced in order to keep the on-shell action finite. Finally, here $\varepsilon^{\mu\nu}=\frac{1}{\sqrt{-g}}\epsilon^{\mu\nu}$ and $K$ is the trace of the extrinsic curvature.

Here, we introduce $L_{CS}= \Phi\varepsilon^{\mu\nu}F_{\mu\nu} $ as the Chern-Simons density (CSd) term associated with the 2D gravity model (\ref{action}).  As our analysis reveals, the CSd term for the wormhole phase is non-zero as one approaches the boundary of the space time. On the other hand, it vanishes asymptotically for the black hole phase. 

This altogether makes a crucial difference between the boundary Free energy densities of these two phases. It is noteworthy to mention that the above is an ``on-shell'' result and does not depend on the choice of the coupling constant $a_2$.

The variation of the action (\ref{action}) yields the following set of equations 
\begin{align}
\Phi &: (R+2)+2a_1\Phi F^2+a_2\varepsilon^{\mu\nu}F_{\mu\nu}=0, \label{eom1}\\
\chi &:\Box\chi=0,\label{eom2}\\
A_{\mu} &: \bigtriangledown _{\mu}\big[2a_1\Phi^{2} F^{\mu\nu}+a_2\Phi\varepsilon^{\mu\nu}\big]=0,\label{eom3}\\
g_{\mu\nu} &:\Box\Phi g_{\mu\nu}-\bigtriangledown_{\mu}\bigtriangledown_{\nu}\Phi-g_{\mu\nu}\Phi+<T_{\mu\nu}>=0,\label{eom4}
\end{align} 
where $<T_{\mu\nu}>$ is the full stress-energy tensor\footnote{We define the stress tensor as $T_{\mu\nu}=\frac{1}{\sqrt{-g}}\frac{\delta S}{\delta g^{\mu\nu}}$.} combining gauge fields and the scalar field
\begin{align}
    <T_{\mu\nu}>\hspace{1mm}&=\hspace{1mm}<T^{(gauge)}_{\mu\nu}>+<T^{(\chi)}_{\mu\nu}>,\label{eom4a}\\
    <T^{(gauge)}_{\mu\nu}>\hspace{1mm}&=a_1\Phi^{2}\Big[2g^{\alpha\beta}F_{\mu\alpha}F_{\nu\beta}-\frac{1}{2}F^2g_{\mu\nu}\Big],\label{eom4b}\\
    <T^{(\chi)}_{\mu\nu}>\hspace{1mm}&=(\partial_{\mu}\chi)(\partial_{\nu}\chi)-\frac{1}{2}g_{\mu\nu}(\partial \chi)^2.\label{eom4c}
    \end{align}
\subsection{Euclidean wormholes}  \label{secwh}
We begin by considering the possibilities for a charged Euclidean wormhole solution of ($\ref{action}$) whose geometry is described by a double trumpet \cite{Garcia-Garcia:2020ttf} having two asymptotics. The corresponding space-time metric is expressed as
\begin{align}\label{gauge}
    &ds^2=\frac{1}{\cos^2\rho}(d\tau^2+d\rho^2)\hspace{1mm};\hspace{2mm}-\frac{\pi}{2}\leq\rho\leq\frac{\pi}{2}\hspace{1.5mm},\hspace{1.5mm} \tau\sim\tau+b,
\end{align}
where $b$ is the periodicity of the Euclidean time $(\tau)$.

Given (\ref{gauge}), we solve the equations of the motion (\ref{eom1})-(\ref{eom4}) using the static gauge
\begin{align}\label{gauge1}
A_{\tau}=\xi(\rho),\hspace{1mm}A_{\rho}=0,\hspace{1mm}\Phi=\Phi(\rho)\hspace{1mm},\hspace{1mm}\chi=\chi(\rho).
\end{align}

Using (\ref{gauge}) and (\ref{gauge1}), it is now trivial to find the solution for the scalar field
\begin{align}\label{chisol}
    \chi=C_1\rho+C_2,
\end{align}
where $C_1$ and $C_2$ are the integration constants. 

On the other hand, the equations of motion for the dilaton (\ref{eom1}) and the gauge field (\ref{eom3}) can be reduced down to a single equation of the form
\begin{align}\label{constraint}
    \Phi\cos^2\rho(\partial_{\rho}\xi)=\frac{a_2}{2a_1},
\end{align}
where we set the integration constant to zero for the consistency of the wormhole solution. 
\subsubsection{Stress-energy tensor}
Our next task would be to compute the full stress-energy tensor $<T_{\mu\nu}>$ combining both the $U(1)$ gauge fields $(A_{\mu})$ as well as the scalar field $(\chi)$
\begin{align}\label{stfull}
    <T_{\mu\nu}>\hspace{1mm}&=\hspace{1mm}<T^{(gauge)}_{\mu\nu}>+<T^{(\chi)}_{\mu\nu}>,
\end{align}
where $<T^{(\chi)}_{\mu\nu}>$ denotes the stress-energy tensor for the scalar field $(\chi)$ and $<T^{(gauge)}_{\mu\nu}>$ corresponds to the stress-energy tensor for the $U(1)$ gauge fields $(A_{\mu})$.

The general strategy would be to break the stress-energy tensor (\ref{stfull}) into classical and quantum pieces as follows
\begin{align}\label{breakref2}
    <T_{\mu\nu}>_{cl}\hspace{1mm}&=\hspace{1mm}<T_{\mu\nu}^{(\chi)}>_{cl}\hspace{1mm}+\hspace{1mm}<T_{\mu\nu}^{(gauge)}>_{cl},\nonumber\\
     <T_{\mu\nu}>_{qm}\hspace{1mm}&=\hspace{1mm}<T_{\mu\nu}^{(\chi)}>_{qm}\hspace{1mm}+\hspace{1mm}<T_{\mu\nu}^{(gauge)}>_{qm},
\end{align}
where $`cl$'stands for the classical and $`qm$' stands for the quantum stress-energy tensor.

The classical part of the stress-energy tensor can be computed directly substituting (\ref{gauge})-(\ref{chisol}) into (\ref{eom4b})-(\ref{eom4c}). On the other hand, we use the point-splitting method of \cite{book} in order to fix the quantum stress-energy tensor ($ <T_{\mu\nu}>_{qm} $).

 The classical and the quantum part of the scalar stress-energy tensor ($ <T^{(\chi)}_{\mu\nu}>_{qm} $) is quite straightforward to obtain. These results can be summarised as follows \cite{Garcia-Garcia:2020ttf}
 \begin{align}
     <T^{(\chi)}_{\rho\rho}>_{cl}\hspace{1mm}&=- <T^{(\chi)}_{\tau\tau}>_{cl}\hspace{1mm}=\frac{C_1^2}{2},\label{sst1}\\
      <T_{\rho\rho}^{(\chi)}>_{qm}&=\frac{1}{24\pi}-\frac{1}{24\pi\cos^2\rho}+X_{\rho\rho}(b),\nonumber\\
      X_{\rho\rho}(b)&=-\sum_{p\in \mathbb{Z}}\frac{p\pi}{b^2\tanh(2\pi^2p/b)},\label{sst2}\\
      <T_{\tau\tau }^{(\chi)}>_{qm}&=-\frac{1}{24\pi}-\frac{1}{24\pi\cos^2\rho}+X_{\tau\tau}(b),\nonumber\\
      X_{\tau\tau}(b)&=\sum_{p\in \mathbb{Z}}\frac{p\pi}{b^2\tanh(2\pi^2p/b)}.\label{sst3}
 \end{align}

The major part of the present computation therefore involves estimating the quantum stress-energy tensor for the $U(1)$ gauge fields. The classical part of the stress-energy tensor can be obtained using (\ref{eom4b}) and (\ref{gauge})
\begin{align}\label{classst}
    <T_{\rho\rho}^{(gauge)}>_{cl}\hspace{1mm}=\hspace{1mm}<T_{\tau\tau}^{(gauge)}>_{cl}\hspace{1mm}=\hspace{1mm}a_1\Phi^{2}\cos^2\rho(\partial_{\rho}\xi)^2.
\end{align}

The derivation of the quantum stress-energy tensor ($ <T_{\mu\nu}^{(gauge)}>_{qm} $) for the $U(1)$ gauge fields (\ref{breakref2}) is discussed in detail in the Appendix \ref{QSTref2}. These results are obtained in a straightforward way by considering a double trumpet geometry (\ref{gauge}). 

Below, we summarise these results as
 \begin{align}
      <T_{\tau\tau +}^{(gauge)}>_{qm}\hspace{1mm}=\hspace{1mm}<T_{\rho\rho +}^{(gauge)}>_{qm}\hspace{1mm}=&\hspace{1mm}\frac{1}{\mathcal{A}_4(\rho)}\Bigg[a_1\exp\big(p_0(\pi-2\rho)/2\big)\cos^2\rho\Phi^2\Big(4\mathcal{A}_1(\rho)+\nonumber\\
      &(\pi-2\rho)\big(\mathcal{A}_2(\rho)-\mathcal{A}_3(\rho)\big)\Big)\Bigg],\label{stgaugepref2}\\
 <T_{\tau\tau -}^{(gauge)}>_{qm}\hspace{1mm}=\hspace{1mm}<T_{\rho\rho -}^{(gauge)}>_{qm}\hspace{1mm}=&\hspace{1mm}\frac{1}{\mathcal{B}_3(\rho)}\Bigg[a_1\exp(p_0\rho)\cos^2\rho\Phi^2\Big(\mathcal{B}_1(\rho)+\nonumber\\&(\pi+2\rho)\mathcal{B}_2(\rho)\Big)\Bigg],\label{stgaugemref2}
 \end{align}
 where the details of the functions $\mathcal{A}_1(\rho),.. \mathcal{A}_4(\rho), \mathcal{B}_1(\rho),.. \mathcal{B}_3(\rho)$ are given in the Appendix \ref{QSTref2} and the subscripts $`\pm$' denote the  expectation values near the boundary limits, $\rho=\pm\frac{\pi}{2}$.

Substituting (\ref{sst1})-(\ref{sst3}), (\ref{classst}), (\ref{stgaugepref2}) and (\ref{stgaugemref2}) into (\ref{stfull}) one finally obtains the full stress-energy tensor combining the $U(1)$ gauge field and the scalar field $(\chi)$. The complete stress-energy tensor (\ref{stfull}) is then used in the next section to calculate the dilaton $(\Phi)$ profile in the asymptotic limits $(\rho\rightarrow\pm\frac{\pi}{2})$ of the wormhole space time.
  \subsubsection{Solving for $\Phi$ and $\xi$}\label{secsolphi}
  Before we proceed further, let us first calculate the boundary stress-energy tensor\footnote{A similar calculation is also discussed in \cite{Rathi:2021aaw}. } using the  Gibbons-Hawking-York term (\ref{action})
  \begin{align}\label{boundary}
   T_{\tau\tau}\Big|_{boundary}=\frac{1}{\sqrt{-\gamma}}\frac{\delta S_{GHY}}{\delta \gamma ^{\tau\tau}}\hspace{1mm},\hspace{1mm}\text{where}\hspace{2mm}S_{GHY}=\int d \tau \sqrt{-\gamma}\Phi2K,
  \end{align}
  which plays a crucial role in obtaining the near boundary profile for the dilaton ($\Phi$).
  
  On solving (\ref{boundary}) using the double trumpet geometry (\ref{gauge}) one finds
  \begin{align}
      T_{\tau\tau}\Big|_{+}=\alpha\Phi\sec{\rho}\tan{\rho}\hspace{2mm}\text{and}\hspace{2mm}T_{\tau\tau}\Big|_{-}=\beta\Phi\sec{\rho}\tan{\rho},
  \end{align}
  where the subscripts $\pm$ denote the stress-energy tensors near the asymptotic limits $\rho_{L,R}\sim\pm\frac{\pi}{2}$ with $\alpha,\beta$ being constants.
  
 Given the double trumpet geometry (\ref{gauge}), the equation of motion (\ref{eom4}) for the metric turns out to be
  \begin{align}\label{phi1}
      \partial_{\rho}^2\Phi-\tan{\rho} \partial_{\rho}\Phi-\frac{\Phi}{\cos^2{\rho}}+<T_{\tau\tau}>=0,
  \end{align}
  where $<T_{\tau\tau}>\hspace{1mm}=\hspace{1mm}<T_{\tau\tau}^{(gauge)}>+<T_{\tau\tau}^{(scalar)}>+<T_{\tau\tau}^{(boundary)}>$. 
  
  Upon solving (\ref{phi1}), one finds the following ``asymptotic'' profiles for the dilaton\footnote{The functions $ F_{\pm} $ play crucial role while determining the Free energy (density) near the boundary of the wormhole space time. This is due to the fact these functions appear explicitly in the asymptotic profiles for the dilaton ($ \Phi_{\pm} $) which in turn carry information about the Free energy (density) near the asymptotic boundary of the wormhole spacetime. These constants in $ F_{\pm} $ are further constrained by the fact that the left and the right temperatures ($ T^{(wh)}_{\pm} $) of the wormhole solution must be identified. This further reduces the number of independent constants to $ r_0 $ and $ \alpha $ which finally appear in the expression for the Free energy (\ref{fedp1}). These constants can be further replaced in terms of the chemical potential ($ \mu $) (see (\ref{finalmu}) ) and the coupling constant(s) which eventually removes all the ambiguities in the expression of the Free energy (\ref{freenergy}).}
  \begin{align}\label{phisol}
    \Phi\Big|_{\rho\rightarrow\frac{\pi}{2}} = \Phi_+=\frac{-F_+}{\big(\frac{\pi}{2}-\rho\big)}-\frac{1}{24\pi}+\alpha r_0\hspace{2mm}\text{and}\hspace{2mm}\Phi\Big|_{\rho\rightarrow-\frac{\pi}{2}}=\Phi_-=\frac{F_-}{\big(\rho+\frac{\pi}{2}\big)}-\frac{1}{24\pi}-\beta s_0,
  \end{align}
  where the details are given in the Appendix \ref{expref2} where the subscripts $`\pm$' denote the leading order terms in $\Phi$.

Using (\ref{constraint}) and (\ref{phisol}), the asymptotic profiles for the gauge field turn out to be
 \begin{align}\label{gaugesol1}
           \xi_{\pm}=\frac{a_2}{2a_1 F_{\pm}}\log\Big(\pi\mp 2\rho\Big)+\mu_{\pm},\hspace{1mm}
       \end{align}
       where $\mu_{\pm}$ denote the chemical potentials near the boundaries $\rho_{L,R}\sim\pm\frac{\pi}{2}$
       \begin{align}\label{gaugesol2}
            \mu_{+}=\frac{a_2}{2a_1}\frac{1}{F_{+}}\Bigg(-1-\frac{192F_{+}^2\pi^2}{(1-24\pi r_0\alpha)^2}\Bigg)\log(-48F_+\pi),\\
            \mu_{-}=\frac{a_2}{2a_1}\frac{1}{F_{-}}\Bigg(-1-\frac{192F_{-}^2\pi^2}{(1+24\pi s_0\beta)^2}\Bigg)\log(-48F_-\pi).
       \end{align}      
       
     At this stage, it is important to notice that the associated Chern-Simons density (CSd) takes a finite value in the asymptotic limits
       \begin{align}\label{ref2cswh}
           L_{CS}^{(wh)}\Big|_{\rho\rightarrow\pm\frac{\pi}{2}}=-2\cos^2\rho\Phi\partial_{\rho}\xi=-\frac{a_2}{a_1},
       \end{align}
  where the on-shell condition (\ref{constraint}) is imposed. Therefore, it plays a significant role while estimating the Free energy (density) for the boundary theory.
 \subsection{Euclidean black holes}\label{secebh}
 We now move on towards constructing the Euclidean black hole solution of (\ref{action}). We further use these solutions to discuss the thermal properties of the 2D black hole. 

We solve these equations (\ref{eom1})-(\ref{eom4}) ``perturbatively'' treating the couplings $a_1$ and $a_2$ as expansion parameters. We express these background fields using the static gauge 
\begin{align}\label{bhansatzref2}
    ds^2&=e^{2\omega(z)}(d\tau^2+dz^2)\hspace{1mm},\nonumber\\
    A_{\mu}&\equiv(A_{\tau}(z),0)\hspace{1mm},\hspace{1mm}\Phi=\Phi(z)\hspace{1mm},\hspace{1mm}\chi=\chi(z).
\end{align}

Next, we expand these background fields schematically as \cite{Lala:2020lge}
\begin{align}\label{fexp}
    \mathcal{H}= \mathcal{H}_0+a_1\mathcal{H}_1+a_2\mathcal{H}_2\hspace{1mm},\hspace{2mm}|a_1|<<1\hspace{1mm},\hspace{1mm}|a_2|<<1,
\end{align}
where $\mathcal{H}$ stands for any of the fields $\omega,\Phi,A_{\tau}$ and $\chi$. 

Here, the subscript $`0$' stands for the pure JT gravity solution while the other two subscripts $`1$'and $`2$' denote the associated corrections due to the $U(1)$ gauge fields \cite{Lala:2020lge}. We discuss all these in detail in the Appendix \ref{bhsolref2}. 

A straightforward computation the on-shell Chern-Simons density (CSd) for the black hole phase shows 
\begin{align}\label{ref2bhcsdm}
  L_{CS}^{bh}\Big|_{z\rightarrow 0}=-2 e^{-2\omega_0}\Phi_0\partial_zA_{\tau}\Big|_{z\rightarrow 0}=-4 b_3z\Big|_{z\rightarrow 0} =0,
\end{align}
where $r=\sqrt{r_H}\coth(2\sqrt{r_H}z)$.  Therefore, unlike the wormhole phase, its contribution can be ignored while obtaining various thermodynamic entities (like Free energy density for example) near the boundary of the black hole spacetime. 

Finally, the space-time metric for the Euclidean black hole (\ref{bhansatzref2}) turns out to be
\begin{align}\label{stm}
    ds^2\approx4(r^2-r_H)(1+2a_1\omega_1)\Bigg(d\tau^2+\frac{dr^2}{4(r_H-r^2)^2}\Bigg),
\end{align}
where $\omega_1$ is given in Appendix \ref{bhsolref2} and the black hole horizon is located at $r=\sqrt{r_H}$.
\section{Thermodynamics}\label{thermoref2}
 We now examine the thermal properties of the wormhole and the black hole solutions those were obtained previously in Section \ref{grsetup}. From the periodicity of the Euclidean time $(\tau)$, one can identify the temperature ($T$) associated with these solutions. 
 
Finally, we estimate the Free energy density ($\mathcal{F}$) and the total charge ($ Q $) and express them as a function of temperature ($T$) and the chemical potential ($\mu$). In other words, we treat both the temperature $(T)$ and the chemical potential $(\mu)$ as independent thermodynamic variables where one of them can be tuned while keeping the other fixed.
 \subsection{Wormholes}
 \subsubsection{Temperature}\label{tempsecref2}
In order to determine the temperature of the wormhole, we impose the following boundary conditions\footnote{These boundary conditions simply follow from the asymptotic structures of the dilaton ($\Phi$) (\ref{phisol}) and the space-time metric ($g_{\mu\nu}$) (\ref{gauge}), where $\epsilon$ is the UV cutoff. } \cite{Maldacena:2018lmt}-\cite{Garcia-Garcia:2020ttf} on $\Phi$ and $g_{\mu\nu}$
 \begin{align}\label{bc1}
     \Phi\sim\frac{\phi}{\epsilon}\hspace{2mm}\text{and}\hspace{2mm}ds^2\Big|_{\rho\rightarrow\pm\frac{\pi}{2}}\sim\frac{du_{\pm}^2}{\epsilon^2},
 \end{align}
 where $u_{\pm}$ are identical to $\tau$ and with the periodicity conditions $u_{\pm}\sim u_{\pm}+\phi\beta_{\pm}$. Here,  $\beta_{\pm}$ correspond to the inverse temperatures near the asymptotics $\rho_{L,R}=\pm\frac{\pi}{2}$.
 
Finally, using (\ref{gauge}), (\ref{phisol}) and (\ref{bc1}), we identify the temperature associated with the wormhole solution near the asymptotics as 
 \begin{align}\label{npbcond1}
     T^{(wh)}_{\pm}=\mp\hspace{1mm}\frac{F_{\pm}}{b}, 
 \end{align}
 where the functions $F_{\pm}$ are given in the Appendix \ref{expref2}.
 
  Notice that, the right temperature $(T^{(wh)}_+)$ of the wormhole near $\rho_R\sim\frac{\pi}{2}$ is different from that of its left temperature $(T^{(wh)}_-)$ near $\rho_L=-\frac{\pi}{2}$. However, setting $\beta=-\alpha,\hspace{1mm}C_3=C_4=\eta,\hspace{1mm}q_0=-p_0,\hspace{1mm}B(-\pi/2,m)=-B(\pi/2,m),\hspace{1mm}D(-\pi/2,m)=-D(\pi/2,m),\hspace{1mm}s_0=r_0,\hspace{1mm}s_1=-r_1,\hspace{1mm}s_2=r_2,\hspace{1mm}s_3=-r_3,$ and $\mu_+=-\mu_-=\mu$, we find that $T^{(wh)}_+= T^{(wh)}_-=T_{(wh)}$.

The identification of the chemical potentials ($ \mu_{\pm} $) reveals an useful identity of the form
\begin{align}\label{finalmu}
      \mu = \frac{a_2}{2a_1bT_{(wh)}}\Bigg(1+\frac{192b^2\pi^2T_{(wh)}^2}{(1-24\pi r_0\alpha)^2}\Bigg)\log(48\pi b T_{(wh)}),
  \end{align}
  which is further used in order to remove ambiguities in the Free energy density.
  \subsubsection{Free energy }
 The Free energy is defined using the Euclidean path integral\footnote{An exact computation of the Free energy would indeed require the computation of the full bulk integral first and thereby taking its asymptotic limits ($ \rho \rightarrow \pm \frac{\pi}{2} $) for some fixed radial coordinate that approaches the boundary. Taking the boundary limit is important because the dual field theory we conjecture about is supposed live on this boundary. Ideally, this should be conjectured as the Free energy density of the dual field theory. However, as far as the present computation is concerned, this turns out to be a quite non-trivial task due to the complicated profile of the dilaton ($ \Phi $) which appears to be an important element of the bulk integral. Therefore, to deal with the situation, one has to \emph{approximate} the integral by considering its limiting value near the boundary of the spacetime. In other words, the ``boundary'' Free energy that is estimated in this paper, is defined as the integral that is evaluated using the \emph{asymptotic} data of the bulk fields where we ignore some of the IR degrees of freedom those might come from the interior of the bulk.}
  \begin{align}\label{fwh}
      F_{(wh)}=-\beta^{-1}\log Z_E^{(wh)},\hspace{2mm}Z_E^{(wh)}=e^{-S_{(wh)}^{(os)}},
  \end{align}
  where $Z_E$ is the Euclidean partition function and $S_{(wh)}^{(os)}$ stands for the Euclidean on-shell action corresponding to the wormhole solutions (\ref{phisol}) and (\ref{gaugesol1}).
  
Recall that, in Section \ref{secsolphi}, we determine asymptotic profile (\ref{phisol}) for the dilaton ($\Phi$). These asymptotic data are used to calculate the boundary Free energy density (\ref{action}) associated with the wormhole phase.
  
The regularised Free energy density $(\mathcal{F}^{(reg)}_{(wh)})$ of the boundary theory is defined through the following integral 
  \begin{align}\label{fedwh}
      F_{(wh)}^{(reg)}\Big|_{\rho\rightarrow\pm\frac{\pi}{2}}=\int d\tau\sqrt{-\gamma}\mathcal{F}_{(wh)}^{(reg)},
  \end{align}
 where the ``regularised'' Free energy density is expressed as\footnote{Here the divergent piece is absorbed using the following counter term $S_{ct}=\int d\tau\sqrt{-\gamma}\big(2\frac{F_+}{\delta}\big)$, where $\delta$ being the UV cutoff.}
  
       \begin{align}\label{fedp1}
      \mathcal{F}_{(wh)}^{(reg)}=&\hspace{1mm}T_{(wh)}\Bigg(-\frac{2a_1b^2T_{(wh)}^2(1-24\pi r_0\alpha)^4\mu^2}{(192b^2\pi^2T_{(wh)}^2+(1-24\pi r_0\alpha)^2)^2\log(48b\pi T_{(wh)})^2}-\frac{1}{12\pi}+2r_0\alpha\Bigg).
      \end{align}
  
  Using (\ref{finalmu}), the above expression (\ref{fedp1}) further simplifies as
  \begin{eqnarray}
  \label{freenergy}
\mathcal{F}_{(wh)}^{(reg)}=  \hspace{1mm}T_{(wh)}\Bigg(-\frac{a_2^2}{2a_1}-\frac{1}{12\pi}\sqrt{\frac{192a_2\pi^2b^2T_{(wh)}^2\log(48\pi b T_{(wh)})}{2a_1b T_{(wh)} \mu-a_2\log(48\pi b T_{(wh)})}}\Bigg).
  \end{eqnarray}
 \subsubsection{Charge} 
 The derivation of the $ U(1) $ charge ($ Q $) follows from the definition of the $ U(1) $ current\footnote{Our analysis follows closely the algorithm developed by authors in \cite{Castro:2008ms}. } \cite{Castro:2008ms} 
 \begin{align}\label{chargedef}
     J^{\mu}=\frac{1}{\sqrt{-g}}\frac{\delta S_{JT}}{\delta A_{\mu}}.
 \end{align}
 
 The variation of (\ref{action}) with respect to the gauge field ($A_{\mu}$) yields
 \begin{align}\label{chargevar}
     \delta S_{JT}=&\int d^2x\sqrt{-g}\nabla_{\mu}\Big(-4 a_1\Phi^2F^{\mu\nu}-2a_2\Phi \varepsilon^{\mu\nu}\Big)\delta A_{\nu} \nonumber\\
     &+ \int d\tau\sqrt{-\gamma} n_{\rho}\Big(4 a_1\Phi^2F^{\rho\tau}+2a_2\Phi \varepsilon^{\rho\tau}\Big)\delta A_{\tau},
 \end{align}
 where $n_{\rho}$ is the unit normal vector to the boundary ($\rho=\pm \frac{\pi}{2}$). Notice that, the first term on the R.H.S. in (\ref{chargevar}) vanishes on-shell. In other words, one is only left with an integral that is evaluated in the asymptotic limit(s) where the dual field theory is living.
 
Using the on-shell condition in (\ref{chargevar}), we finally obtain the boundary current as 
\begin{align}\label{chargectbdy}
    J^{\tau}_{bdy}=n_{\rho}\Big(4 a_1\Phi^2F^{\rho\tau}+2a_2\Phi \varepsilon^{\rho\tau}\Big),
\end{align}
where the subscript $``bdy"$ denotes the current evaluated near the boundary. 

Finally, the $ U(1) $ charge associated with the wormhole phase is define as\footnote{We conjecture this as the global $ U(1) $ charge associated with the dual field theory living on the boundary of the wormhole spacetime.}
\begin{align}\label{bdycharge}
    Q_{(wh)}=\int d\tau \sqrt{-\gamma}J^{\tau}_{bdy}.
\end{align}

Using the asymptotic data (\ref{phisol}) and (\ref{gaugesol1}) together with (\ref{finalmu}), we finally obtain
\begin{align}
\label{boundarycharge}
    Q_{(wh)}= \frac{a_2}{12\pi T_{(wh)}}\sqrt{\frac{192a_2\pi^2b^2T_{(wh)}^2\log(48\pi b T_{(wh)})}{2a_1b T_{(wh)} \mu-a_2\log(48\pi b T_{(wh)})}}.
\end{align}
 
Like Free energy density (\ref{freenergy}), the boundary $ U(1) $ charge (\ref{boundarycharge}) is also free from ambiguities and is fixed by the coupling constant ($ a_2 $) and the periodicity ($ b $) in the Euclidean time. Here, it is noteworthy to mention that both the regularised Free energy density $ (\mathcal{F}_{(wh)}^{(reg)})$ and the total charge ($ Q_{(wh)}$) of the wormhole solution are equal on both the boundaries $\rho_{L,R}$. This follows using  the relations between the constants as mentioned in Section \ref{tempsecref2}.  
\subsection{Black holes}
We now compute the Free energy density and the total charge associated with the black hole solution those are obtained previously in Section \ref{secebh}. The basic philosophy and the physical considerations behind these derivations are the same as those for the wormholes which we therefore prefer not to repeat here.

To begin with, we compute the Hawking temperature\footnote{One can obtain the same expression (\ref{ht}) using the periodicity arguments of the Euclidean time $(\tau)$ in an expansion near the horizon.} ($T_H$) of the 2D black hole \cite{Rathi:2021aaw} 
\begin{align}\label{ht}
    T_H=\frac{1}{2\pi}\sqrt{\frac{1}{4}g^{\tau\tau}g^{rr}(\partial_{r}g_{\tau\tau})^2}\Bigg|_{r\rightarrow\sqrt{r_H}}=\frac{\sqrt{r_H}}{\pi}.
\end{align}
  
  The regularised Free energy density $(\mathcal{F}^{(reg)}_{(bh)})$ for the boundary theory is defined through the following integral
  \begin{align}\label{fedwh}
      F_{(bh)}^{(reg)}\Big|_{r\rightarrow\infty}=\int d\tau\sqrt{-\gamma}\mathcal{F}_{(bh)}^{(reg)},
  \end{align}
   where the ``regularised'' Free energy density is expressed as\footnote{Here the divergences are absorbed using the counter term, $S_{ct}=-\int d\tau\sqrt{-\gamma}\big(2+4a_1b_5+2a_1b_9)r$. The origin of these integration constants are shown in detail in the Appendix \ref{bhsolref2}.}
\begin{align}
\label{bhfreeenery}
    \mathcal{F}_{(bh)}^{(reg)}=-\frac{a_1}{2}T_{H}\Bigg(T_{H}+\frac{d_1}{T_{H}^2}+d_2\Bigg).
\end{align}

Here, we rescale the Free energy density by a constant $d_0=-\pi^2b_5+2\pi(b_4+b_5)$ and the other constants can be expressed in terms of $d_0$ as $d_1=-\frac{b_1b_6}{\pi^2d_0}$ and $d_2=-\frac{4b_8}{d_0}$.

The (regularised) boundary charge for the black hole phase can be obtained in a similar way as in the case for wormholes. After a suitable rescaling by $ b_3 $, this yields 
\begin{align}
\label{c2}
  Q_{(bh)}\Big|_{z=0}=\frac{4a_1}{T_{H}},
\end{align}
which finally depends only on the coupling constant $ a_1 (>0)$ of the theory.
\section{Phase transition}\label{secphasetrans}
Finally, with all these solutions in hand, we are now in a position to explore the thermal stability of our solutions with respect to the temperature ($ T $). The key observables in this regard are the regularised Free energy densities (\ref{freenergy}), (\ref{bhfreeenery}) and the global charges (\ref{boundarycharge}), (\ref{c2}) those were obtained previously in Section \ref{thermoref2}. 

\begin{figure}
\includegraphics[scale=.30]{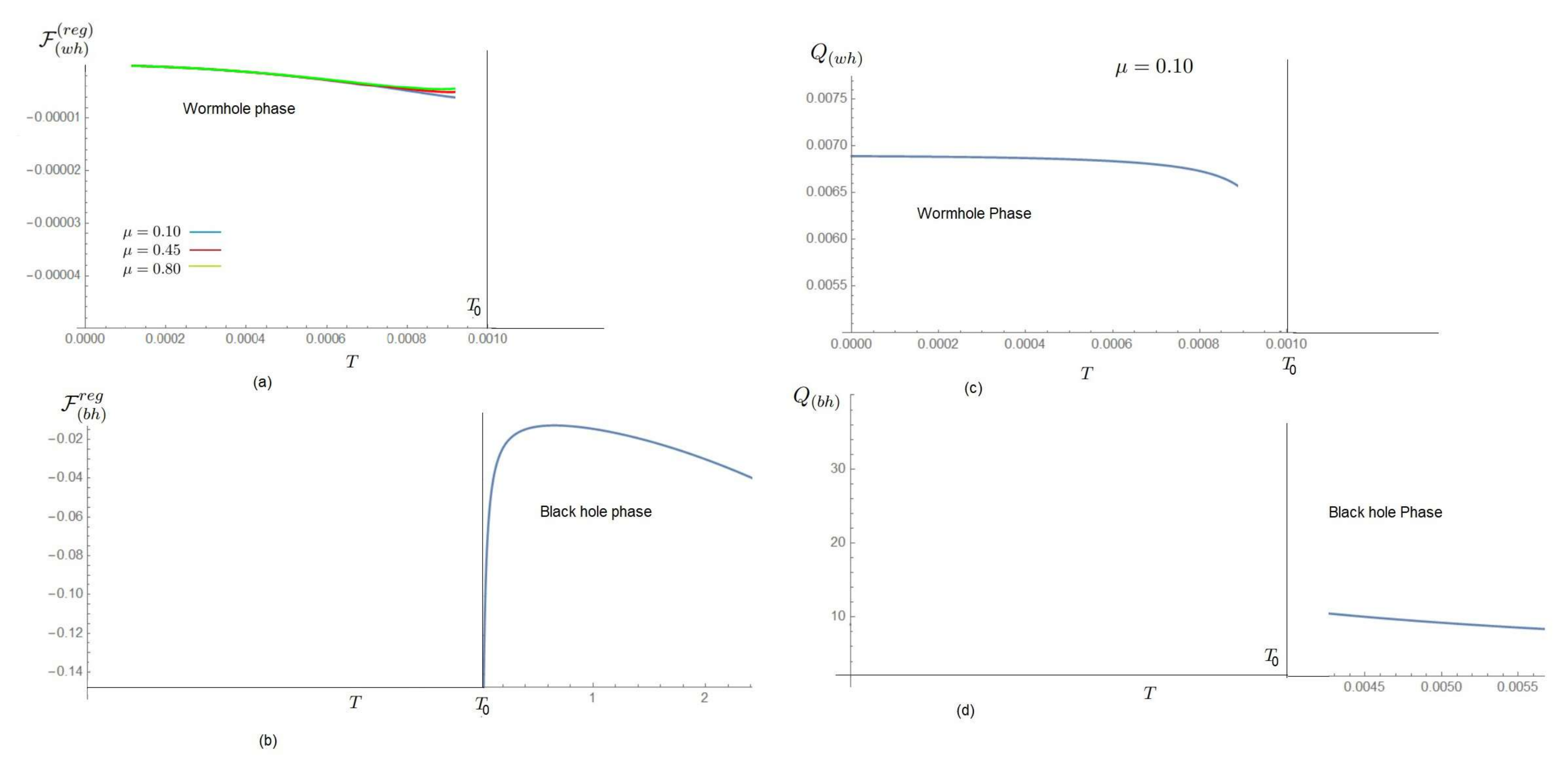}
\caption{Figures (a) and (b) represent the variations of the boundary Free energy densities $(\mathcal{F})$ with temperature ($T$) and at a fixed chemical potential $(\mu)$. On the other hand, the figures (c) and (d) illustrate the variation of the total charge $(Q)$ with temperature ($T$). Here we denote both the wormhole and the black hole temperatures as $T$. We identify the critical temperature as $T_0=\frac{1}{48b\pi}$, where $ b $ is the periodicity of the Euclidean time. Here, we set the basic parameters of the theory as $b=6.63145,a_1=0.009,a_2=0.0009$. These parameters are chosen in order to achieve a best fit for the plot. The choices for the coupling constants ($ a_1 $, $ a_2 \ll 1$) are always less than one in a perturbative expansion (see (\ref{fexp})). Finally, we set the remaining constants as, $d_1=1$ and $d_2=1$.} 
\label{figref2four}
\end{figure}

The  variations of the ``boundary'' Free energy densities ((\ref{freenergy}) and (\ref{bhfreeenery})) are shown in the figures \ref{figref2four}(a) and \ref{figref2four}(b). As these figures reveal, for sufficiently low temperatures $(T<<T_0)$, the regularised Free energy $(\mathcal{F}_{(wh)}^{reg})$ of the charged wormhole  solution remains as a constant indicating the presence of a ``gapped'' phase in the dual (conjectured) two-site complex SYK model at finite chemical potential $(\mu=\mu_0)$.

As the temperature of the system is further increased, we observe a ``discontinuous'' change (see figure \ref{figref2four}(a)) at $T= T_0$, characterising the onset of a \emph{first} order phase transition. On increasing the temperature beyond $T=T_0$, the wormhole phase becomes unstable and passes over to an Euclidean (two) black hole system as shown in figure \ref{figref2four}(b). The dual counterpart of this phase is conjectured to be a ``hot'' complex SYK model at finite chemical potential ($ \mu $). 

We also plot the total charge ($Q$) of the system as a function of temperature ($T$) (see figures \ref{figref2four}(c) and \ref{figref2four}(d)). Notice that, as we move towards the critical temperature ($T= T_0$), we observe a discontinuous jump in the total charge ($Q$) of the system which is quite reminiscent to that of \cite{Garcia-Garcia:2020vyr}. However, unlike in \cite{Garcia-Garcia:2020vyr}, here one should conjecture about a two-site ``uncoupled'' complex SYK rather than a coupled one. 

\section{A qualitative discussion on the conjectured SYK dual }\label{npbnewsecconject}
In this Section, we qualitatively argue the structure of the dual field theory corresponding to the JT gravity set up (\ref{action}). The notable features of this set up is two fold (i) the presence of $U(1)$ gauge field (which sources a chemical potential ($\mu$) for the dual SYK) and (ii) the boundary contributions coming from the bulk Chern-Simons term. The Chern-Simons term plays a significant role while obtaining the wormhole phase in the dual gravity description which is therefore expected to play a vital role while constructing the low temperature phase of the dual SYK model.

Notice that, the present work corresponds to the phase stability of (charged) Euclidean wormhole solutions and not  ``traversable wormhole'' solutions of \cite{Maldacena:2018lmt},\cite{Garcia-Garcia:2020vyr}. Therefore, it is expected that the field theory (that is dual to the gravitational set up (\ref{action})) should represent a two-site \emph{uncoupled} complex SYK model in the presence of a global $U(1)$ symmetry \cite{Garcia-Garcia:2020vyr},\cite{Zhang:2020szi},\cite{Davison:2016ngz}-\cite{Bulycheva:2017uqj}.

However, unlike the previous examples  \cite{Garcia-Garcia:2020vyr},\cite{Zhang:2020szi},\cite{Davison:2016ngz}-\cite{Bulycheva:2017uqj}, the conjectured dual Hamiltonian must contain an additional contribution ($H_{CS}$) due to the bulk Chern-Simons term in (\ref{action}). The dual Hamiltonian could be schematically expressed as
\begin{align}\label{neweomsyk}
    \tilde{H}_{SYK}=\sum_{m=1,2}H_{SYK}^{(m)}+H_{CS},
\end{align}
where $H_{SYK}^{(m)}$ , {($m=1,2$)} is the usual complex (uncoupled) Hamiltonian where the superscript $m$ denotes the number of copies of the SYK model \cite{Garcia-Garcia:2020vyr},\cite{Zhang:2020szi},\cite{Davison:2016ngz}-\cite{Bulycheva:2017uqj}.

Given the above facts, we conjecture that there exists an Euclidean action corresponding to the dual Hamiltonian (\ref{neweomsyk}) which can be schematically expressed as
\begin{align}\label{npbactioncsyk1}
    S=\int d\tau\Bigg[\frac{1}{2}\psi^{\dag}_i(\partial_{\tau}-\mu)\psi_i-\tilde{H}_{SYK}\Bigg],
\end{align}
and should be thought of as a straightforward generalization of \cite{Gaikwad:2018dfc} in the presence of Chern-Simons contributions. 

The boundary contribution due to the bulk Chern-Simons term can be estimated by expanding the bulk action (\ref{action}) in the near boundary limit which turns out to be 

\begin{align}\label{npbactioncsyk2}
    S_{CS}\sim a_2\int d\tau\mu\Phi_b,
\end{align}
where $\Phi_b$ stands for the boundary value of the bulk dilaton ($\Phi$) that acts as a coupling constant for the dual SYK model under consideration. 

Notice that, here we perform the  bulk calculations using a static gauge (\ref{gauge1}) and therefore $\Phi_b$ does not explicitly depend on the Euclidean time ($\tau$). This suggests that, for the present model, the Chern-Simons contribution (\ref{npbactioncsyk2}) acts as a constant shift to the boundary Hamiltonian (\ref{npbactioncsyk1}).

Now, one can further investigate various thermodynamic properties and in particular the Free energy associated with the (two-site) complex SYK model (\ref{neweomsyk}) and compare it with the Free energy calculations in the dual gravitational description. For this purpose, one requires to define the statistical average of the path integral and the Green's function in the Large N limit associated with the complex SYK model (\ref{neweomsyk}). The next step would be to solve the corresponding Schwinger-Dyson\footnote{Alternatively, one can also compute the Free energy of the system by exact diagonalization technique of the Hamiltonian \cite{Garcia-Garcia:2020ttf}. This method is quite useful over the Schwinger-Dyson approach whenever there is no inter-site coupling between the two copies of the SYK model.} (SD) equations for Green's function and calculate the grand canonical potential of the system \cite{Garcia-Garcia:2020vyr},\cite{Garcia-Garcia:2022xsh}-\cite{Garcia-Garcia:2022adg}. Using the grand potential, one can further estimate the Free energy pertinent to the dual model (\ref{neweomsyk}).  

It is expected that the above Free energy calculation in its infrared limit (or low energy regime) would match with the Free energy of the 2D Einstein-Maxwell-dilaton gravity system (\ref{action}) in the regime of small temperature ($T<<1$) and charge ($Q<<1$) where the chemical potential ($\mu$) is held fixed and set to be small (see Figures \ref{figref2four}(a) and \ref{figref2four}(c)). In other words, the bulk wormhole solutions at low temperature ($T<<1$) and low charge densities ($Q<<1$) should represent the low energy phase of the dual SYK model (\ref{neweomsyk}).

In connection to the above, it is also expected that the Free energy computed on the dual SYK side should exhibit a zero slope (flat region) in its infrared which is similar to that of the JT gravity set up in the regime of small temperature ($T$) and chemical potential ($\mu$) (see Figure \ref{figref2four}(a)). This would indicate the possibilities for finding a ``wormhole phase'' on the SYK counterpart of the duality. It would be indeed an interesting project to explore all the above directions in the near future.

\section{Concluding remarks}\label{secconc}
We conclude by highlighting the key results of the paper and add some further remarks on various interesting future extensions of the present work.  

The present paper is an attempt to understand the phases of Euclidean wormhole solutions in the presence of an abelian one form. In other words, the current analysis is a generalisation of the gravitational sector of \cite{Garcia-Garcia:2020ttf} in the presence of a finite charge density and/or chemical potential ($\mu$). The presence of the chemical potential ($\mu$) eventually reveals a richer phase structure which we summarise below.

At low temperatures, the gravitational sector of the system turns out to be a charged wormhole solution. As the temperature is increased beyond $T=T_0$, the wormhole phase undergoes a first order phase transition and transits into a two black hole system at finite charge density. Like boundary Free energy (density), the boundary $ U(1) $ charge $(Q)$ also undergoes a discontinuous change \cite{Garcia-Garcia:2020vyr} at $ T=T_0 $. 

Below we outline several interesting future extensions of the present work.

$\bullet$ It would be indeed an interesting project to look for the boundary 1D theory for our model and identify the corresponding Schwarzian degrees of freedom. The Schwarzian action can be obtained following a standard procedure in which one integrates out the dilaton ($ \Phi $). In the process, one is left with the boundary term which is regarded as the low energy effective theory living on the boundary. Clearly, this theory should contain an additional parameter in it, which is the chemical potential ($ \mu $) of the boundary field theory.

Below, we outline a toy model calculation, in which we show schematically how does this extra term is generated due to the presence of the chemical potential ($ \mu $). To begin with, we consider an $ AdS_2 $ metric of the form
 \begin{equation}
     ds^2=\frac{1}{z^2}(dt^2+dz^2),
 \end{equation}
 where the boundary is located at, $z \sim \varepsilon \sim 0$.
 
Taking the first term on the R.H.S. of (\ref{action}) into account, one can show that the boundary action typically looks like (here we absorb the coupling constants of the bulk theory into $ \mu $)
\begin{eqnarray}
S_B \sim \int du~\Phi_r (u)  \text{Sch}\lbrace t(u), u\rbrace +\mu  \int du \frac{\Phi_r^2 (u)}{\varepsilon},
\end{eqnarray}
where we parameterize the coordinates ($t , z$) in terms of the boundary time coordinate ($u$) and $\Phi_r (u)$ is the coupling constant of the boundary field theory which is defined in terms of the boundary data of the form $\Phi|_{bdy}=\frac{\Phi_r}{\epsilon}$ \cite{Maldacena:2016upp}. It would be indeed an interesting project to explore this action in detail an extract Green's functions etc. out of it. We leave all these issues for future investigations.

$\bullet$ Recall that, in Section \ref{secwh} we compute the ``annealed'' Free energy \cite{Garcia-Garcia:2020ttf} of the wormhole solution and explore the associated phase stability (see Section \ref{secphasetrans}). However, one can further refine the computation by including the corrections due to replica wormholes \cite{Engelhardt:2020qpv,Almheiri:2019qdq} and compute the ``quenched'' Free energy \cite{Garcia-Garcia:2020ttf}. Our next goal would be to explore the properties of ``quenched'' Free energy and study the corresponding phase stability of the configuration at finite chemical potential \cite{HR}.

$\bullet$ It would be interesting to generalise the present calculation in the presence of $SU(2)$ Yang-Mills fields and study the phase stability of the configuration. However, it has been found in \cite{Lala:2020lge} that $SU(2)$ Yang-Mills fields are responsible for the Hawking-Page transition in 2D gravity models. Therefore, we expect a much richer phase structure in this model.

$\bullet$ Finally, one can further generalise the present model in the presence of the higher derivative interactions \cite{Rathi:2021aaw} and look for their imprints on the corresponding phase stability.
\section*{Acknowledgments}
The authors are indebted to the authorities of Indian Institute of Technology, Roorkee for their unconditional support towards researches in basic sciences. DR would like to acknowledge The Royal Society, UK for financial assistance. DR  would also like to acknowledge the Grant (No. SRG/2020/000088) received from The Science and Engineering Research Board (SERB), India.
\appendix
\section{3D to 2D reduction}
\label{dimreduction}
We begin by considering $ AdS_3 $ gravity coupled to Maxwell Chern-Simons term of topological gauge theories \cite{Deser:1981wh}-\cite{VanMechelen:2019ebr}
  \begin{align}\label{3daction}
      S_{(3)}=\int d^3x\sqrt{-g_{(3)}}\Bigg(R_{(3)}+2+a_1F_{MN}F^{MN}+a_2\epsilon^{MNP}A_M\partial_N A_P+2(\partial_M f)^2\Bigg),
  \end{align}
 where $R_{(3)}$ is the 3D Ricci scalar, $f$ is the scalar field and $\epsilon^{MNP}=\frac{\varepsilon^{MNP}}{\sqrt{-g_{(3)}}}$ is the Levi-Civita tensor. Here $(M,N)$ represent the 3 dimensional indices.
 
In order to obtain the desired JT gravity model (\ref{action}), we propose a reduction ansatz of the following form 
 \begin{align}\label{3dansatz}
     ds_{(3)}^2&=\Phi(x^{\mu})^{-1}ds_{(2)}^2+\Phi(x^{\mu})^{2}dz^2,\hspace{2mm}ds_{(2)}^2=\Tilde{g}_{\mu\nu}(x^{\alpha})dx^{\mu}dx^{\nu},\nonumber\\
    A_{\mu}&\equiv A_{\mu}(x^{\nu}),\hspace{1mm}A_z\equiv \kappa(x^{\mu}),\hspace{1mm}f\equiv f(x^{\mu}),
 \end{align}
where $(\mu,\nu)$ are the 2 dimensional indices and $z$ is the compact dimension.

Substituting  (\ref{3dansatz}) into (\ref{3daction}) and integrating over compact dimension, we find
\begin{align}\label{2dactionreduce}
    S_{(2)}=&\int d^2x\sqrt{-g_{(2)}}\Bigg(\Phi R_{(2)}+2\big(1+(\partial_{\mu}f)^2\Phi\big)-\Phi^{-1}(\partial_{\mu}\Phi)^2+a_1\Phi^2F^2+\nonumber\\
    &2a_1\Phi^{-1}(\partial_{\mu}\kappa)^2 +a_2\epsilon^{\mu\nu}_{(2)}F_{\mu\nu}\kappa\Bigg).
\end{align}

Next, we redefine the fields as
\begin{align}\label{redefine}
 \kappa (\rho)=\frac{\Phi}{\sqrt{2a_1}},\hspace{2mm}f(\rho)=\int d\rho\sqrt{\frac{\Tilde{g}_{\rho\rho}(\Phi-1)}{\Phi}},
\end{align}
where $\rho$ is the radial direction of $ AdS_2 $.

Finally, substituting (\ref{redefine}) into (\ref{2dactionreduce}), we obtain the 2D gravity model
\begin{align}
    S_{(2)}=\int d^2x\sqrt{-g_{(2)}}\Bigg(\Phi \big(R_{(2)}+2\big)+a_1\Phi^2F^2+\Tilde{a}_2\epsilon^{\mu\nu}_{(2)}F_{\mu\nu}\Phi\Bigg),
\end{align}
where we define $\Tilde{a}_2=\frac{a_2}{\sqrt{2a_1}}$.
\section{Quantum stress-energy tensor for gauge fields }\label{QSTref2}
In this Appendix, we compute the expectation value for the quantum stress-energy tensor \cite{Garcia-Garcia:2020ttf} in the double trumpet background (\ref{gauge}). We use the point-splitting method of \cite{book}. As mentioned previously, the following derivation is technically different from that of the scalar field ($\chi$) \cite{Garcia-Garcia:2020ttf}. The reason for this stems from the fact that gauge fields in 2D are non-conformal. Therefore, one has to carry out a first principle derivation of the quantum stress-energy tensor considering the double trumpet geometry as the background.

The equation of motion for the gauge field, $A_{\mu}$ (\ref{eom3}) in the double trumpet background (\ref{gauge}) turns out to be 
\begin{align}\label{cylgaugeeom}
   \hat{\textbf{F}}A_{\tau}=J(\rho),
\end{align}
where $\hat{\textbf{F}}=\partial^2_{\rho}+(f(\rho)-2\tan(\rho))\partial_{\rho}$ is the corresponding differential operator. 

Furthermore, here we define
$$f(\rho)=2\Phi^{-1}\partial_{\rho}\Phi\hspace{1mm},\hspace{2mm} J(\rho)=\frac{ a_2}{2a_1}\frac{\partial_{\rho}\Phi}{\Phi^{2}\cos^2\rho}.$$

In order to proceed further, we segregate out gauge field components from (\ref{action})
\begin{align}\label{gaugeaction}
    S_{gauge}=a_1\int_{(I)} d^2x\sqrt{-g}\Phi^{2}F^2+a_2\int_{(II)} d^2x\sqrt{-g}\Phi\varepsilon^{\mu\nu}F_{\mu\nu}.
\end{align}

Notice that, the variation of the second integral $(II)$ with respect to the metric $g_{\mu\nu}$ vanishes. Therefore, only the first integral $(I)$ contributes to the expectation value of the quantum stress-energy tensor for gauge fields (\ref{breakref2}). 

In order to take the variation of (\ref{gaugeaction}), we decompose the (first) integral $(I)$ into the bulk and the boundary pieces as follows
\begin{align}\label{gaugeactionsplit}
    S_{gauge}&=2a_1\int d^2xA_{\nu}\partial_{\mu}\Big[\sqrt{-g}\Phi^{2}K^{\mu\alpha\nu\beta}\partial_{\beta}A_{\alpha}\Big]+a_2\int d^2x\sqrt{-g}\Phi\varepsilon^{\mu\nu}F_{\mu\nu}+S_{boundary},\nonumber\\
    K^{\mu\alpha\nu\beta}&=g^{\mu\alpha}g^{\nu\beta}-g^{\mu\beta}g^{\nu\alpha}.
\end{align}

The variation of (\ref{gaugeactionsplit}) with respect to the bulk metric $g^{\eta\kappa}$  yields the following expression
\begin{align}\label{gaugevariation1}
    &\frac{\delta S_{gauge}}{\delta g^{\eta\kappa}}=2a_1\int d^2x\Bigg[A_{\nu}\Big[\Phi^{2}\Big\{-\frac{1}{2}\Big(\partial_{\mu}(g_{\eta\kappa}\sqrt{-g})+\sqrt{-g}g_{\lambda\sigma}\frac{\delta(\partial_{\mu}g^{\lambda\sigma})}{\delta g^{\eta \kappa}}\Big)K^{\mu\alpha\nu\beta}-\nonumber\\
    &\frac{1}{2}\sqrt{-g}g_{\eta\kappa}(\partial_{\mu}K^{\mu\alpha\nu\beta})+\sqrt{-g}\Big(g^{\nu\beta}\frac{\delta(\partial_{\mu}g^{\mu\alpha})}{\delta g^{\eta\kappa}}+g^{\mu\alpha}\frac{\delta(\partial_{\mu}g^{\nu\beta})}{\delta g^{\eta\kappa}}-g^{\nu\alpha}\frac{\delta(\partial_{\mu}g^{\mu\beta})}{\delta g^{\eta\kappa}}-g^{\mu\beta}\frac{\delta(\partial_{\mu} g^{\nu\alpha})}{\delta g^{\eta\kappa}}\Big)\Big\}\Big]\dot\nonumber\\
    &(\partial_{\beta}A_{\alpha})+\Phi^{2}\Big\{A_{\nu}(\partial_{\eta}\sqrt{-g})g^{\nu\beta}\partial_{\beta}A_{\kappa}+A_{\eta}(\partial_{\mu}\sqrt{-g})g^{\mu\alpha}\partial_{\kappa}A_{\alpha}-A_{\nu}(\partial_{\eta}\sqrt{-g})g^{\nu\beta}\partial_{\kappa}A_{\beta}-\nonumber\\
    &A_{\eta}(\partial_{\mu}\sqrt{-g})g^{\mu\beta}\partial_{\beta}A_{\kappa}+\sqrt{-g}\Big(A_{\nu}\partial_{\eta}(g^{\nu\beta})\partial_{\beta}A_{\kappa}+A_{\eta}\partial_{\mu}(g^{\mu\alpha})\partial_{\kappa}A_{\alpha}-A_{\nu}\partial_{\eta}(g^{\nu\alpha})\partial_{\kappa}A_{\alpha}-\nonumber\\
    &A_{\eta}\partial_{\mu}(g^{\mu\beta})\partial_{\beta}A_{\kappa}\Big)\Big\}+\sqrt{-g}\Big\{A_{\nu}g^{\nu\beta}\partial_{\eta}(\Phi^{2}\partial_{\beta}A_{\kappa})-\frac{1}{2}A_{\nu}g_{\eta\kappa}K^{\mu\alpha\nu\beta}\partial_{\mu}(\Phi^{2}\partial_{\beta}A_{\alpha})+\nonumber\\
    &A_{\eta}g^{\mu\alpha}\partial_{\mu}(\Phi^{2}\partial_{\kappa}A_{\alpha})-A_{\nu}g^{\nu\alpha}\partial_{\eta}(\Phi^{2}\partial_{\kappa} A_{\alpha})-A_{\eta}g^{\mu\beta}\partial_{\mu}(\Phi^{2}\partial_{\beta}A_{\kappa})\Big\}\Bigg].
\end{align}

In the double trumpet background (\ref{gauge}), the above expression (\ref{gaugevariation1}) further simplifies as
\begin{align}\label{gaugevariation2}
    \frac{\delta S_{gauge}}{\delta g^{\tau\tau}}= \frac{\delta S_{gauge}}{\delta g^{\rho\rho}}=-a_1\int d^2x\sqrt{-g}A_{\tau}\hat{\textbf{L}} A_{\tau},
\end{align}
where we ignore all the derivatives of the metric variation i.e. $ \Big|\frac{\partial_{\mu}(\delta g^{\mu\alpha})}{\delta g^{\eta\kappa}}\Big|<<1$ and  $\hat{\textbf{L}}$ is the differential operator such that $\hat{\textbf{L}}A_{\tau}=\cos^2\rho\partial_{\rho}(\Phi^{2}\partial_{\rho}A_{\tau})$.  

Using (\ref{gaugevariation2}) as well as the point splitting method \cite{Garcia-Garcia:2020ttf}, the expectation value for the quantum stress-energy tensor\footnote{Notice that, there are two possible ways in which the differential operator $\hat{\textbf{L}}$ can act on the Green's function $G(\rho,\tau;\rho',\tau')$. However, both the possibilities give the same result. Therefore, we consider the average of both the possibilities in the definition of the expectation value of the stress tensor.} turns out to be 
\begin{align}\label{stcylgauge}
    <T_{\tau\tau }^{(gauge)}>_{qm}\hspace{1mm}=\hspace{1mm}<T_{\rho\rho}^{(gauge)}>_{qm}\hspace{1mm}= \frac{a_1}{2}\Big[&\lim_{x'\rightarrow x}\cos^2\rho\partial_{\rho}\Big(\Phi^{2}(\rho')\partial_{\rho'}G(\rho,\tau;\rho',\tau')\Big)+\nonumber\\
    &\lim_{x\rightarrow x'}\cos^2\rho'\partial_{\rho'}\Big(\Phi^{2}(\rho)\partial_{\rho}G(\rho,\tau;\rho',\tau')\Big)\Big],
\end{align}
where $x\equiv(\rho,\tau)$,  $x'\equiv(\rho',\tau')$.

Here $G(\rho,\tau;\rho',\tau')$ is the Green's function which satisfies the following equation
\begin{align}\label{green1}
   \hat{\textbf{F}}G(\rho,\tau;\rho',\tau')=-\delta(\rho-\rho')\delta(\tau-\tau'), 
\end{align}
where the operator $\hat{\textbf{F}}$ is given by (\ref{cylgaugeeom}).

In order to solve (\ref{green1}), we consider the following Fourier decomposition of $G(\rho,\tau;\rho',\tau')$ and $\delta(\tau-\tau')$
\begin{align}
    G(\rho,\tau;\rho',\tau')&=\sum_{m,m'\in \mathbb{Z}}\Tilde{G}(\rho,m;\rho',m')e^{2\pi i (m\tau+m'\tau')/b}\label{fd1},\\
    \delta(\tau-\tau')&=\frac{1}{b}\sum_{m,m'\in \mathbb{Z}}e^{2\pi i (m\tau+m'\tau')/b}\delta_{m+m'}.\label{fd2}
\end{align}

Plugging (\ref{fd1}) and (\ref{fd2}) into (\ref{green1}), we obtain
\begin{align}\label{green2}
    \Big[\partial^2_{\rho}+(f(\rho)-2\tan(\rho))\partial_{\rho}\Big]\Tilde{G}(\rho,m;\rho',m')=-\frac{1}{b}\delta(\rho-\rho')\delta_{m+m'}.
\end{align}

Now, we solve (\ref{green2}) using the following properties: \\

   $\bullet$ $\Tilde{G}$ is continuous in the limit $\rho\rightarrow \rho'$ : $\Tilde{G}(\rho)=\Tilde{G}(\rho')$.
   
   $\bullet$ The derivative of $\Tilde{G}$ is discontinuous in the limit\footnote{Here, we integrate (\ref{green2}) with respect to $\rho$ from $\rho=\rho'-\epsilon$ to $\rho=\rho'+\epsilon$ and take the limit $\epsilon\rightarrow 0.$} $\epsilon\rightarrow 0$ namely,
   \begin{align}\label{disc}
       \frac{d \Tilde{G}}{d\rho}\Big|_{\rho'+\epsilon}-\frac{d \Tilde{G}}{d\rho}\Big|_{\rho'-\epsilon}=-\frac{1}{b}.
   \end{align}

Notice that, $f(\rho)$ in (\ref{green2}) is an unknown function of $\rho$. Therefore one cannot solve (\ref{green2}) exactly for all values of $\rho$. However, we are interested in the near boundary analysis, therefore we solve (\ref{green2}) in the near boundary limits, $\rho\rightarrow\pm\frac{\pi}{2}$.\\

$\bullet$ \underline{Case 1} : Near  $\rho\sim\rho_R\sim\frac{\pi}{2} $ and $-\frac{\pi}{2}<\rho'<\rho<\frac{\pi}{2}$

Setting $m'=-m$ and upon solving equation (\ref{green2}) near $\rho\sim\frac{\pi}{2}$, we obtain 
 \begin{align}\label{npbgp1}
     \Tilde{G}_+(\rho,\rho',m)= &\hspace{1mm} \frac{1}{4}A(\rho',m)\Bigg(\frac{2\exp(p_0(\pi-2\rho)/2)}{\pi-2\rho}-p_0\text{Ei}(p_0(\pi-2\rho)/2)\Bigg)\nonumber \\
     &+B(\rho',m),
 \end{align}\\
 where $p_0=f\Big(\frac{\pi}{2}\Big)$ and the subscript $`+$' denotes the value of Green's function near the right boundary $\rho_R\sim+\frac{\pi}{2}$.\\
 
 $\bullet$ \underline{Case 2} : Near $\rho\sim\rho_L\sim-\frac{\pi}{2}$ and $-\frac{\pi}{2}<\rho<\rho'<\frac{\pi}{2}$

 On setting $m'=-m$ and solving equation (\ref{green2}) near $\rho\sim-\frac{\pi}{2}$, we obtain
 \begin{align}\label{npbgp2}
      \Tilde{G}_-(\rho,\rho',m)= &\hspace{1mm}  \frac{1}{4}C(\rho',m)\Bigg(-\frac{2\exp(-q_0(\pi+2\rho)/2)}{\pi+2\rho}-q_0\text{Ei}(-q_0(\pi+2\rho)/2)\Bigg)\nonumber \\
     &+D(\rho',m),
 \end{align}
  where $q_0=f\Big(-\frac{\pi}{2}\Big)$ and the subscript $`-$' denotes the value of Green's function near the left boundary $\rho_L\sim-\frac{\pi}{2}$.
  
Notice that, the differential equation (\ref{green2}) contains the derivative of $\Tilde{G}(\rho,m;\rho',m')$ with respect to the quantity ``$\rho$''. Therefore, the entities $A(\rho',m)$, $B(\rho',m)$, $C(\rho',m)$ and $D(\rho',m)$ in (\ref{npbgp1}) and (\ref{npbgp2}) appears as an integration constants. However, one can compute the functions $A(\rho',m)$ and $C(\rho',m)$ using the properties of the Green's function those were mentioned in (\ref{disc}) but the functions $B(\rho',m)$ and $D(\rho',m)$ still remain undetermined. Therefore, we perform all the analysis keeping the general form of the functions $B(\rho',m)$ and $D(\rho',m)$ and finally impose a suitable condition on these functions (\ref{npbcond1}) using the fact that the left temperature of the wormhole near $\rho_L\sim-\frac{\pi}{2}$ must be identified with the right temperature near $\rho_R\sim\frac{\pi}{2}$.

Finally, (\ref{fd1}) can be systematically expressed as 
 \begin{align}\label{greenfunc}
     G(\rho,\tau;\rho',\tau')= \left\{
 \begin{array}{lr} 
      \sum\limits_{m\in \mathbb{Z}} \Tilde{G}_+(\rho,\rho',m)e^{2\pi i (\tau-\tau')m/b} & -\frac{\pi}{2}<\rho'<\rho<\frac{\pi}{2} \\
     \sum\limits_{m\in \mathbb{Z}} \Tilde{G}_-(\rho,\rho',m)e^{2\pi i (\tau-\tau')m/b}  & -\frac{\pi}{2}<\rho<\rho'<\frac{\pi}{2} 
      \end{array}
      \right.,
 \end{align}
 where the function $A(\rho',m)$ and $C(\rho',m)$ could be expressed as
 \begin{align}
    &A(\rho',m)=-\frac{1}{b}(\pi-2\rho')^2\Big(4b B(\rho',m)-4b D(\rho',m)+(\pi+2\rho')\big(2+\nonumber\\
     &\exp(q_0(\pi+2\rho')/2)q_0(\pi+2\rho')\text{Ei}(-q_0(\pi+2\rho')/2)\big)\Big)\Big(-p_0(\pi-2\rho')^2\times\nonumber\\
     &\text{Ei}(p_0(\pi-2\rho')/2)+\exp(p_0(\pi-2\rho')/2)\big(4\pi+\exp(q_0(\pi+2\rho')/2)q_0(\pi+2\rho')^2\times\nonumber\\
     &\text{Ei}(-q_0(\pi+2\rho')/2)\big)\Big)^{-1},\\
    &C(\rho',m)=\frac{1}{b}\exp(\pi q_0/2+(p_0+q_0)\rho')\Big(2\exp(p_0(\pi-2\rho')/2)(\pi+2\rho')^2(\pi-2\rho'-\nonumber\\
    &2b B(\rho',m)+2b D(\rho',m))-p_0(\pi^2-4\rho'^2)^2\text{Ei}(p_0(\pi-2\rho')/2)\Big)\Big(4\exp(p_0\pi/2)\pi-\nonumber\\
    &\exp(p_0\rho')p_0(\pi-2\rho')^2\text{Ei}(p_0(\pi-2\rho')/2)+\exp(\pi(p_0+q_0)/2+q_0\rho')q_0(\pi+2\rho')^2\times\nonumber\\
    &\text{Ei}(-q_0(\pi+2\rho')/2)\Big)^{-1}.
 \end{align}

Finally, plugging (\ref{greenfunc}) into (\ref{stcylgauge}), we obtain 
\begin{align}
      <T_{\tau\tau +}^{(gauge)}>_{qm}\hspace{1mm}=\hspace{1mm}<T_{\rho\rho +}^{(gauge)}>_{qm}\hspace{1mm}=&\hspace{1mm}\frac{1}{\mathcal{A}_4(\rho)}\Bigg[a_1\exp\big(p_0(\pi-2\rho)/2\big)\cos^2\rho\Phi^2\Big(4\mathcal{A}_1(\rho)+\nonumber\\
      &(\pi-2\rho)\big(\mathcal{A}_2(\rho)-\mathcal{A}_3(\rho)\big)\Big)\Bigg],\label{stgaugepref2new}\\
       <T_{\tau\tau -}^{(gauge)}>_{qm}\hspace{1mm}=\hspace{1mm}<T_{\rho\rho -}^{(gauge)}>_{qm}\hspace{1mm}=&\hspace{1mm}\frac{1}{\mathcal{B}_3(\rho)}\Bigg[a_1\exp(p_0\rho)\cos^2\rho\Phi^2\Big(\mathcal{B}_1(\rho)+\nonumber\\
       &\hspace{1mm}(\pi+2\rho)\mathcal{B}_2(\rho)\Big)\Bigg],\label{stgaugemref2new}
 \end{align}
where the functions $\mathcal{A}_1(\rho),.. \mathcal{A}_4(\rho), \mathcal{B}_1(\rho),.. \mathcal{B}_3(\rho)$ are given by
\begin{align}
    &\mathcal{A}_1(\rho)=\Bigg(4b B(\rho,m)-4b D(\rho,m)+(\pi+2\rho)\Big(2+\exp(q_0(\pi+2\rho)/2)q_0(\pi+2\rho)\times\nonumber\\
    &\text{Ei}(-q_0(\pi+2\rho)/2)\Big)\Bigg)\Bigg(-p_0(\pi-2\rho)^2\text{Ei}(p_0(\pi-2\rho)/2)+\exp(p_0(\pi-2\rho)/2)\times\nonumber\\
    &\Big(4\pi+\exp(q_0(\pi+2\rho)/2)q_0(\pi+2\rho)^2\text{Ei}(-q_0(\pi+2\rho)/2)\Big)\Bigg),\\
   & \mathcal{A}_2(\rho)=\Bigg(4b B(\rho,m)-4b D(\rho,m)+(\pi+2\rho)\Big(2+\exp(q_0(\pi+2\rho)/2)q_0(\pi+2\rho)\times\nonumber\\
    &\text{Ei}(-q_0(\pi+2\rho)/2)\Big)\Bigg)\Bigg(2\exp(p_0(\pi-2\rho)/2)p_0(\pi-2\rho)+4p_0(\pi-2\rho)\times\nonumber\\
   &\text{Ei}(p_0(\pi-2\rho)/2)-\exp(p_0(\pi-2\rho)/2)p_0\Big(4\pi+\exp(q_0(\pi+2\rho)/2)q_0(\pi+2\rho)^2\times\nonumber\\
   &\text{Ei}(-q_0(\pi+2\rho)/2)\Big)+\exp(p_0(\pi-2\rho)/2)q_0(\pi+2\rho)\Big(2+\exp(q_0(\pi+2\rho)/2)\times\nonumber\\
   &(4+\pi q_0+2q_0\rho)\text{Ei}(-q_0(\pi+2\rho)/2)\Big)\Bigg),\\
   &\mathcal{A}_3(\rho)= \Bigg(-p_0(\pi-2\rho)^2\text{Ei}(p_0(\pi-2\rho)/2)+\exp(p_0(\pi-2\rho)/2)\Big(4\pi+\nonumber\\
   &\exp(q_0(\pi+2\rho)/2)q_0(\pi+2\rho)^2\text{Ei}(-q_0(\pi+2\rho)/2)\Big)\Bigg)\Bigg(4+2\exp(q_0(\pi+2\rho)/2)\times\nonumber\\
   &q_0(\pi+2\rho)\text{Ei}(-q_0(\pi+2\rho)/2)+q_0(\pi+2\rho)\Big(2+\exp(q_0(\pi+2\rho)/2)(2+\pi q_0+\nonumber\\
   &2q_0\rho)\text{Ei}(-q_0(\pi+2\rho)/2)\Big)+4b\partial_{\rho}B(\rho,m)-4b\partial_{\rho}D(\rho,m)\Bigg),\\
  & \mathcal{A}_4(\rho)=b(\pi-2\rho)\Bigg(p_0(\pi-2\rho)^2\text{Ei}(p_0(\pi-2\rho)/2)-\exp(p_0(\pi-2\rho)/2)\Big(4\pi+\nonumber\\
  &\exp(q_0(\pi+2\rho)/2)q_0(\pi+2\rho)^2\text{Ei}(-q_0(\pi+2\rho)/2)\Big)\Bigg)^2,
   \end{align}
    \begin{align}
    &\mathcal{B}_1(\rho)=(p_0+q_0)\Bigg(2\exp(p_0(\pi-2\rho)/2)(\pi+2\rho)^2(\pi-2\rho-2bB(\rho,m)+2bD(\rho,m))\nonumber\\
  &-p_0(\pi^2-4\rho^2)^2\text{Ei}(p_0(\pi-2\rho)/2)\Bigg)\Bigg[\Bigg(4\exp(p_0\pi/2)\pi-\exp(p_0\rho)p_0(\pi-2\rho)^2\times\nonumber\\
  &\text{Ei}(p_0(\pi-2\rho)/2)+\exp\big(\pi(p_0+q_0)/2+q_0\rho\big)q_0(\pi+2\rho)^2\text{Ei}(-q_0(\pi+2\rho)/2)\Bigg)\nonumber\\
  &-\frac{1}{(p_0+q_0)}\Bigg(2\exp(p_0\pi/2)p_0(\pi-2\rho)+2\exp(p_0\pi/2)q_0(\pi+2\rho)+4\exp(p_0\rho)p_0\times\nonumber\\
  &(\pi-2\rho)\text{Ei}(p_0(\pi-2\rho)/2)-\exp(p_0\rho)p_0^2(\pi-2\rho)^2\text{Ei}(p_0(\pi-2\rho)/2)+q_0(\pi+2\rho)\times\nonumber\\
  &4\exp\big((p_0+q_0)\pi/2+q_0\rho\big)\text{Ei}(-q_0(\pi+2\rho)/2)+\exp\big((p_0+q_0)\pi/2+q_0\rho\big)q_0^2(\pi+2\rho)^2\nonumber\\
  &\times\text{Ei}(-q_0(\pi+2\rho)/2)\Bigg)\Bigg],\\
  &\mathcal{B}_2(\rho)=2\Bigg(4\exp(p_0\pi/2)\pi-\exp(p_0\rho)p_0(\pi-2\rho)^2\text{Ei}(p_0(\pi-2\rho)/2)+ q_0(\pi+2\rho)^2\times\nonumber\\
  &\exp\big((p_0+q_0)\pi/2+q_0\rho\big)\text{Ei}(-q_0(\pi+2\rho)/2)\Bigg)\Bigg(\exp(p_0(\pi-2\rho)/2)p_0(\pi-2\rho)(\pi+2\rho)\nonumber\\
  &+4\exp(p_0(\pi-2\rho)/2)(\pi-2\rho-2bB(\rho,m)+2bD(\rho,m))-\exp(p_0(\pi-2\rho)/2)p_0\times\nonumber\\
  &(\pi+2\rho)(\pi-2\rho-2bB(\rho,m)+2bD(\rho,m))+8p_0\rho(\pi-2\rho)\text{Ei}(p_0(\pi-2\rho)/2)-\nonumber\\
  &2\exp(p_0(\pi-2\rho)/2)(\pi+2\rho)(1+b\partial_{\rho}B(\rho,m)-b\partial_{\rho}D(\rho,m))\Bigg),\\
   & \mathcal{B}_3(\rho)=b(\pi+2\rho)^2\Bigg(4\exp(p_0\pi/2)\pi-\exp(p_0\rho)p_0(\pi-2\rho)^2 \text{Ei}(p_0(\pi-2\rho)/2)+\nonumber\\
  &\exp\big((p_0+q_0)\pi/2+q_0\rho\big)q_0(\pi+2\rho)^2  \text{Ei}(-q_0(\pi+2\rho)/2)\Bigg)^2.
\end{align}
Here the subscript $`\pm$' denotes the  expectation value of the quantum stress-energy tensor near the boundary  $\rho=\pm\frac{\pi}{2}$.

The leading order terms in the quantum stress-energy tensor could be expressed as follows
 \begin{align}
      &<T_{\tau\tau +}^{(gauge)}>_{qm}\Big|_{\rho\rightarrow\frac{\pi}{2}}\hspace{1mm}=\hspace{1mm}<T_{\rho\rho +}^{(gauge)}>_{qm}\Big|_{\rho\rightarrow\frac{\pi}{2}}\hspace{1mm}=\hspace{1mm}\frac{a_1}{b \pi}(\pi-2\rho)\Big(\pi+b B(\pi/2,m)-\nonumber\\
       &b D(\pi/2,m)+\exp(\pi q_0)\pi^2q_0\text{Ei}(-\pi q_0)\Big)\Phi\Big(\frac{\pi}{2}\Big)^{2}\Big(1+\exp(\pi q_0)\pi q_0\text{Ei}(-\pi q_0)\Big)^{-1},\label{stcylgaugep}
       \end{align}
       \begin{align}
       &<T_{\tau\tau -}^{(gauge)}>_{qm}\Big|_{\rho\rightarrow-\frac{\pi}{2}}\hspace{1mm}=\hspace{1mm}<T_{\rho\rho -}^{(gauge)}>_{qm}\Big|_{\rho\rightarrow-\frac{\pi}{2}}\hspace{1mm}=\hspace{1mm}\frac{a_1}{b \pi}(\pi+2\rho)\Big(b\exp(p_0\pi)\times \nonumber\\
       &B(-\pi/2,m)-b \exp(p_0\pi)D(-\pi/2,m)+\pi\big(-\exp(p_0\pi)+p_0\pi\text{Ei}(\pi p_0)\big)\Big)\Phi\Big(-\frac{\pi}{2}\Big)^{2}\times\nonumber\\
       &\Big(-\exp(p_0\pi)+\pi p_0\text{Ei}(\pi p_0)\Big)^{-1}.\label{stcylgaugem}
 \end{align}
 
Before we conclude, it is noteworthy to mention that one can also evaluate the expectation value of the quantum stress-energy tensor for the scalar field $\chi$ following the method as described above. In the flat space-time limit ($\rho\rightarrow0$), these results boil down into an expression in the cylindrical coordinates \cite{Garcia-Garcia:2020ttf} as given below 
\begin{align}
     <T_{\rho\rho}^{(\chi)}>_{qm}&\approx\sum_{m\in \mathbb{Z}}-\frac{m\pi}{b^2\tanh(2\pi^2m/b)}+\Bigg(-\frac{m\pi\coth(2\pi^2m/b)}{b^2}-\frac{\tanh(m\pi^2/b)}{4m\pi}\Bigg)\rho^2\nonumber\\
     &=T_{cyl}+O(\rho^2).
\end{align}
\section{Detailed expressions of $F_{\pm}$}\label{expref2}
The functions $F_{\pm}$, are given by 
  \begin{align}
      F_+=&\hspace{1mm}\Bigg(-C_3+\frac{1}{48}\big(\pi(-4r_0+12r_2+\pi(r_1-r_3))-8\big((-1+6\gamma)r_1+r_3\big)\big)\alpha+\nonumber\\
      &r_1\alpha \log\Big(\frac{2}{\pi}\Big)+\frac{\pi}{2}X_{\tau\tau}(b)+\frac{1}{36}\Bigg(-1-9C_1^2\pi+\frac{9a_1(8+\pi^2)}{b \pi}\Big(\pi+b B(\pi/2,m)\nonumber\\
      &-b D(\pi/2,m)+\exp(\pi q_0)\pi^2q_0\text{Ei}(-\pi q_0)\Big)\Phi\Big(\frac{\pi}{2}\Big)^{2}\Big(1+\exp(\pi q_0)\pi q_0\times\nonumber\\
      &\text{Ei}(-\pi q_0)\Big)^{-1}\Bigg)\Bigg),\label{fphi1}\\
        F_-=&\hspace{1mm}\Bigg(\frac{1}{36}+C_4+\frac{C_1^2\pi}{4}+\frac{1}{48}\big(8\big((-1+6\gamma)s_1+s_3\big)+\pi(-4s_0+12s_2+\pi(-s_1+s_3))\big)\beta\nonumber\\
      &+s_1\beta \log\Big(\frac{\pi}{2}\Big)-\frac{\pi}{2}X_{\tau\tau}(b)+\frac{a_1}{4b \pi}(8+\pi^2)\Big(b\exp(p_0\pi) B(-\pi/2,m)-b \exp(p_0\pi)\times\nonumber\\
       &D(-\pi/2,m)+\pi\big(-\exp(p_0\pi)+p_0\pi\text{Ei}(\pi p_0)\big)\Big)\Phi\Big(-\frac{\pi}{2}\Big)^{2}\Big(\exp(p_0\pi)-\nonumber\\
       &\pi p_0\text{Ei}(\pi p_0)\Big)^{-1}\Bigg),\label{fphi2}
       \end{align}
       where we denote $r_0=\hspace{1mm}\Phi\Big |_{\rho=\frac{\pi}{2}},\hspace{1mm}r_1=\Phi'\Big|_{\rho=\frac{\pi}{2}},\hspace{1mm}r_2=\Phi''\Big|_{\rho=\frac{\pi}{2}},\hspace{1mm}r_3=\Phi'''\Big|_{\rho=\frac{\pi}{2}},$ 
       
       \begin{align}
       \hspace{1mm}s_0=\Phi\Big|_{\rho=-\frac{\pi}{2}},\hspace{1mm}s_1=\Phi'\Big|_{\rho=-\frac{\pi}{2}},\hspace{1mm}s_2=\Phi''\Big|_{\rho=-\frac{\pi}{2}},\hspace{1mm}s_3=\Phi'''\Big|_{\rho=-\frac{\pi}{2}}. 
         \end{align}
 Here $\gamma$ is the Euler's constant and $C_3,C_4$ are the integration constants.
\section{Black hole solution}\label{bhsolref2}
In the following section, we evaluate the black hole solution by substituting (\ref{fexp}) into the equations of motion (\ref{eom1})-(\ref{eom4}) and solve them at different orders in the coupling.
\subsection{Zeroth order solutions}
Zeroth order solutions are obtained by setting the expansion parameters as $a_1=a_2=0$. The corresponding equations of motion (\ref{eom1})-(\ref{eom4}) turn out to be
\begin{align}
    \Phi_0''-2e^{2\omega_0}\Phi_0&=0,\label{bh01}\\
    \omega_0''-e^{2\omega_0}&=0,\label{bh02}\\
    \chi_0''&=0.\label{bh03}
\end{align}

On solving (\ref{bh01})-(\ref{bh03}), we find the zeroth order solutions as
\begin{align}\label{bh0sol}
    \omega_0=\frac{1}{2}\log\Bigg(\frac{4r_H}{\sinh^2(2\sqrt{r_H}z)}\Bigg),\hspace{1mm}\Phi_0=\sqrt{r_H}\coth(2\sqrt{r_H}z),\hspace{1mm}\chi_0=b_1z+b_2,
\end{align}
where $r_H,b_1$ and $b_2$ are integration constants. 
\subsection{First order solutions}
We now estimate the leading order contributions due to $U(1)$ gauge fields. The corresponding equations of motion (\ref{eom1})-(\ref{eom4}) turn out to be
\begin{align}
    \Phi_1''-2(\omega_1'\Phi_0'+\omega_0'\Phi_1')-2\chi_0'\chi_1'&=0,\label{bh11}\\
    \omega_1''-2\omega_0''\omega_1-2\Phi_0e^{-2\omega_0}A_{\tau1}'^2&=0,\label{bh12}\\
    \partial_z(\Phi_0^2e^{-2\omega_0}A_{\tau1}')&=0,\label{bh13}\\
    \chi_1''-2\chi_0''\omega_1&=0.\label{bh14}
\end{align}

In order to solve (\ref{bh11})-(\ref{bh14}), we adopt the following change in coordinates 
\begin{align}
    z=\frac{1}{2\sqrt{r_H}}\coth^{-1}\Bigg(\frac{r}{\sqrt{r_H}}\Bigg),
\end{align}
where $r_H$ denotes the location of the black hole horizon.

Upon solving (\ref{bh11})-(\ref{bh14}) we find first order corrections to the background fields as
\begin{align}
A_{\tau1}=&\hspace{1mm}\frac{2b_3}{r}+\mu_{bh},\label{bhsol11}\\
\omega_1=&\hspace{1mm}-\frac{b_3^2}{r_H^2r}\Big(r_H+r^2(-2\log(r)+\log(-\sqrt{r_H}+r)+\log(\sqrt{r_H}+r))\Big)-b_5+\nonumber\\
&\frac{r}{\sqrt{r_H}}\Bigg(b_4+\tanh^{-1}\Bigg(\frac{r}{\sqrt{r_H}}\Bigg)b_5\Bigg),\label{bhsol12}\\
\chi_1=&\hspace{1mm}\frac{b_6}{\sqrt{r_H}}\tanh^{-1}\Bigg(\frac{r}{\sqrt{r_H}}\Bigg)+b_7,\label{bhsol13}\\
    \Phi_1=&\hspace{1mm}\frac{b_3^2}{r_H^2}\Big((-r_H+r^2)(2\log r-\log(-\sqrt{r_H}+r)-\log(\sqrt{r_H}+r))\Big)+\frac{1}{8r_H^{\frac{3}{2}}}\Bigg(8r_H r^2b_4+\nonumber\\
    &4r_H\Bigg(2\sqrt{r_H}r+2r^2\tanh^{-1}\Bigg(\frac{r}{\sqrt{r_H}}\Bigg)+r_H\log(-\sqrt{r_H}+r)-r_H\log(\sqrt{r_H}+r)\Bigg)b_5\nonumber\\
    &+r\Big(-\log(-\sqrt{r_H}+r)+\log(\sqrt{r_H}+r)\Big)b_1b_6\Bigg)+b_8+\rho b_{9}\label{bhsol14},
\end{align}
where $b_3,b_4,.,b_{9}$ are the integration constants and $\mu_{bh}$ is the chemical potential for the black hole phase.

The black hole solution up to leading order in $a_1$ can be summarised as 
\begin{align}
     \omega_{bh}=&\hspace{1mm}\frac{1}{2}\log(-4r_H+4r^2)+a_1\Bigg(
-\frac{b_3^2}{r_H^2r}\Big(r_H+r^2(-2\log(r)+\log(-\sqrt{r_H}+r)+\nonumber\\
&\log(\sqrt{r_H}+r))\Big)-b_5+
\frac{r}{\sqrt{r_H}}\Bigg(b_4+\tanh^{-1}\Bigg(\frac{r}{\sqrt{r_H}}\Bigg)b_5\Bigg)\Bigg),\\
\Phi_{bh}=&\hspace{1mm}r+a_1\Bigg(\frac{b_3^2}{r_H^2}\Big((-r_H+r^2)(2\log r-\log(-\sqrt{r_H}+r)-\log(\sqrt{r_H}+r))\Big)+\nonumber\\
    &\frac{1}{8r_H^{\frac{3}{2}}}\Bigg(8r_H r^2b_4+4r_H\Bigg(2\sqrt{r_H}r+2r^2\tanh^{-1}\Bigg(\frac{r}{\sqrt{r_H}}\Bigg)+r_H\log(-\sqrt{r_H}+r)-\nonumber\\
    &r_H\log(\sqrt{r_H}+r)\Bigg)b_5+r\Big(-\log(-\sqrt{r_H}+r)+\log(\sqrt{r_H}+r)\Big)b_1b_6\Bigg)+b_8+\rho b_{9}\Bigg),\\
    \chi_{bh}=&\hspace{1mm}\frac{b_1\coth^{-1}(r/\sqrt{r_H})}{2\sqrt{r_H}}+b_2+a_1\Bigg(\frac{b_6}{\sqrt{r_H}}\tanh^{-1}\Bigg(\frac{r}{\sqrt{r_H}}\Bigg)+b_7\Bigg),\\
    A_{\tau}^{bh}=&\frac{2b_3}{r}+\mu_{bh}.
\end{align}

\end{document}